\documentclass[english,pccp,preprint,superscriptaddress]{revtex4-1} 
\usepackage{epsfig}
\usepackage{amsmath} 
\usepackage{graphicx} 
\usepackage{longtable} 
\usepackage{hyperref} 
\usepackage{amssymb} 
\usepackage{dcolumn} 
\usepackage{mhchem} 
\usepackage{cleveref} 
\usepackage{subfigure} 
\usepackage{array} 
\usepackage{multirow} 
\usepackage{tabularx} 
\usepackage{cleveref} 
\makeatletter
\renewcommand*{\@fnsymbol}[1]{\ensuremath{\ifcase#1\or \dagger\or *\or \ddagger\or 
   \mathsection\or \mathparagraph\or \|\or **\or \dagger\dagger
   \or \ddagger\ddagger \else\@ctrerr\fi}}
\makeatother

\begin{document} 

\title{Electronic and Optical Properties of the Narrowest Armchair Graphene Nanoribbons Studied by Density Functional Methods} 

\author{Chia-Nan Yeh} 
\thanks{These authors contributed equally to this work.} 
\affiliation{Department of Physics, National Taiwan University, Taipei 10617, Taiwan} 

\author{Pei-Yin Lee} 
\thanks{These authors contributed equally to this work.} 
\affiliation{Department of Physics, National Taiwan University, Taipei 10617, Taiwan} 

\author{Jeng-Da Chai} 
\email[Author to whom correspondence should be addressed. Electronic mail: ]{jdchai@phys.ntu.edu.tw} 
\affiliation{Department of Physics, National Taiwan University, Taipei 10617, Taiwan} 
\affiliation{Center for Theoretical Sciences and Center for Quantum Science and Engineering, National Taiwan University, Taipei 10617, Taiwan} 

\date{\today} 

\begin{abstract} 

In the present study, a series of planar poly(\emph{p}-phenylene) (PPP) oligomers with $n$ phenyl rings ($n = 1 - 20$), designated as $n$-PP, are taken as finite-size models of the 
narrowest armchair graphene nanoribbons with hydrogen passivation. The singlet-triplet energy gap, vertical ionization potential, vertical electron affinity, fundamental gap, optical gap, 
and exciton binding energy of $n$-PP are calculated using Kohn-Sham density functional theory and time-dependent density functional theory with various exchange-correlation density 
functionals. The ground state of $n$-PP is shown to be singlet for all the chain lengths studied. In contrast to the lowest singlet state (i.e., the ground state), the lowest triplet state and 
the ground states of the cation and anion of $n$-PP are found to exhibit some multi-reference character. Overall, the electronic and optical properties of $n$-PP obtained from the 
$\omega$B97 and $\omega$B97X functionals are in excellent agreement with the available experimental data. 

\end{abstract} 

\maketitle

\section{Introduction} 

In recent years, graphene, a single layer of carbon atoms tightly packed into a honeycomb lattice, has received considerable attention due to its remarkable properties and technological 
applications \cite{G1,G2,G3,G4,G5,G6,G7,G8,G9}. Graphene exhibits high carrier mobility and long spin diffusion length, giving promises for graphene-based electronics and spintronics. 
However, as graphene has a vanishing band gap, it cannot be directly adopted for transistor applications. Accordingly, developing methods to open a band gap in graphene is necessary 
for its practical applications. 

To generate a nonvanishing and tunable band gap in graphene, the charge carriers can be confined to quasi-one-dimensional systems, such as graphene nanoribbons (GNRs), long and 
narrow graphene strips. Consequently, several experimental techniques have been developed for synthesizing GNRs \cite{GNRsyn,GNRsyn1,GNRsyn2,GNRsyn3}. Because of their 
fascinating electronic, optical, and magnetic properties, GNRs have recently gained increasing interests \cite{Acene-Carter,GNR7,HM,FG,FG1,FG2,FG3,HM1,HM2,edge,elec,elec1,
Acene-DMRG,Mag,HM3,Acene-Jiang,Acene-RAS-SF,Acene-CCSDT,Acene-2RDM,Acene-CCSDT2,GNRs-CCSDT,TAO-LDA,GNRs-DMRG,GNRs-PHF,GNRs-MRAQCC,TAO-GGA,
GNRs-MRAQCC2,GNRs-TAO,AGNRs-NC,GNR_TDDFT,GNR_DFT1,GNR_DFT2}. However, the electronic and optical properties of GNRs can be very sensitive to their width, length, 
edge shape (zigzag, armchair, or chiral), and edge termination. To properly design GNR-based nanodevices, a thorough understanding of the related parameters governing the electronic 
and optical properties of GNRs is of fundamental and practical significance. 

While there has been a growing interest in GNRs, the study of the electronic and optical properties of long-chain GNRs remains very challenging. From the experimental perspectives, 
the difficulties in the synthesis of long-chain GNRs and their instability following isolation have been attributed to their radical character. Accordingly, there have been very few reported 
experimental data for the properties of long-chain GNRs. From the theoretical perspectives, as GNRs belong to $\pi$-conjugated systems, they may exhibit multi-reference character in 
certain circumstances, where conventional single-reference methods may be inadequate. For instance, zigzag GNRs (ZGNRs), which are GNRs with zigzag shaped edges on both sides, 
have been extensively studied, and long-chain ZGNRs have been found to exhibit polyradical character in their ground states, where the active orbitals are mainly localized at the zigzag 
edges \cite{Acene-DMRG,Acene-Jiang,TAO-LDA,GNRs-DMRG,GNRs-PHF,GNRs-MRAQCC,TAO-GGA,GNRs-MRAQCC2,GNRs-TAO}. 

In contrast to ZGNRs, armchair GNRs (AGNRs), which are GNRs with armchair shaped edges on both sides, are expected to possess relatively large fundamental gaps. However, the 
properties of long-chain AGNRs have not been extensively studied, relative to those of long-chain ZGNRs. We believe that a comprehensive understanding of the properties of AGNRs 
is also essential for the optimal design of GNR-based nanodevices. For a theoretical study of the electronic and optical properties of AGNRs, density functional methods, such as 
Kohn-Sham density functional theory (KS-DFT) \cite{KS1,KS2} (for ground-state properties) and time-dependent density functional theory (TDDFT) \cite{RG} (for excited-state properties), 
are ideal, due to their computational efficiency and reasonable accuracy for large systems \cite{KSrev1,KSrev2,KSrev3,TDrev1}. 

Therefore, in this work, we adopt KS-DFT and TDDFT with various exchange-correlation (XC) density functionals to study the electronic and optical properties of the narrowest AGNRs 
(NAGNRs) with different lengths. The rest of this paper is organized as follows. In Section II, we describe our model systems and computational details. The calculated electronic and 
optical properties are compared with the available experimental data and those obtained from high-level {\it ab initio} methods in Section III. Our conclusions are presented in Section IV.

\section{Model Systems and Computational Details} 

\begin{figure} 
\includegraphics[scale=1.0]{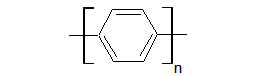} 
\caption{Structure of $n$-PP (C$_{6n}$H$_{4n+2}$), consisting of $n$ benzene rings.} 
\label{fig:PPP} 
\end{figure}

As illustrated in \Cref{fig:PPP}, we adopt a series of planar poly(\emph{p}-phenylene) (PPP) oligomers with $n$ phenyl rings, designated as $n$-PP, as finite-size models of the NAGNRs 
with hydrogen passivation. Note that the number of electrons in $n$-PP (C$_{6n}$H$_{4n+2}$) is $40n + 2$, which rapidly increases with the increase of $n$. Therefore, efficient methods, 
such as KS-DFT and TDDFT, are highly desirable for the study of long-chain $n$-PP. 

For the KS-DFT and TDDFT calculations, we adopt seven XC density functionals, which can be categorized into three different types of density functionals, such as semilocal 
functionals \cite{SF}, global hybrid functionals \cite{hybrid}, and long-range corrected (LC) hybrid functionals \cite{LC1,LC2,LC3,LC4,LC5,LC6,LC7,LC8,LC9,LC10,LC11}: 
\begin{itemize} 
\item semilocal functionals: LDA \cite{LDA1,LDA2}, PBE \cite{PBE}, and BLYP \cite{BLYP1,BLYP2} 
\item global hybrid functionals: PBE0 \cite{PBE0} and B3LYP \cite{B3LYP1,B3LYP2} 
\item LC hybrid functionals: $\omega$B97 \cite{LC6} and $\omega$B97X \cite{LC6} 
\end{itemize} 
for the study of various electronic and optical properties of $n$-PP (up to $n = 20$), involving 
\begin{itemize} 
\item singlet-triplet energy gap 
\item vertical ionization potential 
\item vertical electron affinity 
\item fundamental gap 
\item optical gap 
\item exciton binding energy 
\end{itemize} 
Note that $\omega$B97 and $\omega$B97X have been recently shown to provide excellent performance for a wide range of applications, especially for those closely related to frontier 
orbital energies \cite{LCAC,EB}. 

To estimate the electronic and optical properties of $n$-PP at the polymer limit ($n \to \infty$), a fitting function of the form $(a + b/n)$ is adopted for the extrapolation of the calculated 
and experimental data. Note that this fitting function has been previously adopted to estimate the vertical ionization potential and optical gap of PPP \cite{Eopt3,syn2}. 

In addition, the expectation value of the total spin-squared operator $\langle {\hat{S}}^2 \rangle$ is adopted as a measure of the degree of spin contamination in KS-DFT. For a system 
with strong multi-reference character, the value of $\langle {\hat{S}}^2 \rangle$ obtained from KS-DFT with conventional (semilocal, global hybrid, and LC hybrid) density functionals 
can be significantly different (e.g., more than $10\%$ difference) \cite{tenp} from the exact value $S(S+1)$, where $S$ can be 0 (singlet), 1/2 (doublet), 1 (triplet), 3/2 (quartet), and 
so on. For such a system, KS-DFT employing conventional density functionals can yield unreliable results. To properly describe strong static correlation in such a system, it may be 
essential to adopt multi-reference methods \cite{Acene-DMRG,GNRs-DMRG,GNRs-MRAQCC,GNRs-MRAQCC2} for small-sized systems or 
thermally-assisted-occupation density functional theory (TAO-DFT) \cite{TAO-LDA,TAO-GGA,GNRs-TAO} for medium- to large-sized systems. 

All calculations are performed with a development version of \textsf{Q-Chem 4.0} \cite{Q-Chem}. Results are computed using the 6-31G(d) basis set with the fine grid EML(75,302), 
consisting of 75 Euler-Maclaurin radial grid points \cite{EM} and 302 Lebedev angular grid points \cite{L}. 

As there may be more than one way of calculating the electronic and optical properties using KS-DFT and TDDFT, respectively, we briefly describe how these properties are 
computed as follows.

\subsection{Singlet-Triplet Energy Gap} 

The singlet-triplet energy gap ($E_{\text{ST}}$) of a neutral molecule is defined as 
\begin{equation} 
E_{\text{ST}} = E_{\text{T}} - E_{\text{S}}, 
\label{EST} 
\end{equation} 
the energy difference between the lowest triplet (T) and singlet (S) states, calculated at the respective optimized geometries.

\subsection{Vertical Ionization Potential} 

The vertical ionization potential (IP) of a neutral molecule is defined as 
\begin{equation} 
\text{IP}(1) = E_{\text{total}}(cation) - E_{\text{total}}(neutral), 
\label{IP1} 
\end{equation} 
the energy difference between the cationic and neutral states, calculated at the ground-state geometry of the neutral molecule. 

For the exact KS-DFT, the vertical IP of a neutral molecule is identical to the minus HOMO (highest occupied molecular orbital) energy of the neutral 
molecule \cite{Janak,Fractional,Levy84,1overR,HOMO,HOMO2}, 
\begin{equation} 
\text{IP}(2) = -{\epsilon}_{\text{HOMO}}(neutral). 
\label{IP2} 
\end{equation} 
Accordingly, IP(2) is identical to IP(1) for the exact KS-DFT. For KS-DFT employing approximate XC density functionals, the calculated IP(1) and IP(2) values may differ, 
reflecting the accuracy of the calculated total energies and HOMO energies, respectively.

\subsection{Vertical Electron Affinity} 

The vertical electron affinity (EA) of a neutral molecule is defined as 
\begin{equation} 
\text{EA}(1) = E_{\text{total}}(neutral) - E_{\text{total}}(anion), 
\label{EA1} 
\end{equation} 
the energy difference between the neutral and anionic states, calculated at the ground-state geometry of the neutral molecule. 

By comparing Eq.\ (\ref{IP1}) with Eq.\ (\ref{EA1}), the vertical EA of a neutral molecule is identical to the vertical IP of the anion, which is, for the exact KS-DFT, 
the minus HOMO energy of the anion, calculated at the ground-state geometry of the neutral molecule, 
\begin{equation} 
\text{EA}(2) = -{\epsilon}_{\text{HOMO}}(anion). 
\label{EA2} 
\end{equation} 

In addition, the vertical EA of a neutral molecule is conventionally approximated by the minus LUMO (lowest unoccupied molecular orbital) energy of the neutral molecule, 
\begin{equation} 
\text{EA}(3) = -{\epsilon}_{\text{LUMO}}(neutral). 
\label{EA3} 
\end{equation} 
However, even for the exact KS-DFT, there is a distinct difference between EA(3) and EA(2), due to the derivative discontinuity ($\Delta_{xc}$) of the XC density 
functional \cite{HOMO2,Fractional,DD2,DD3,DD4,DD5,DD6,DD7,DD8,DD9,DD10,DD11}: $\text{EA}(3) - \text{EA}(2) 
= {\epsilon}_{\text{HOMO}}(anion) - {\epsilon}_{\text{LUMO}}(neutral) = \Delta_{xc}$. 
Global and LC hybrid functionals, which belong to the generalized Kohn-Sham (GKS) method \cite{GKS} (not {\it pure} KS-DFT), effectively incorporate a fraction of $\Delta_{xc}$ of 
the XC density functional in KS-DFT. A recent study has shown that the difference between ${\epsilon}_{\text{HOMO}}(anion)$ and ${\epsilon}_{\text{LUMO}}(neutral)$ is small for 
LC hybrid functionals \cite{SC4}. Hence, for LC hybrid functionals, EA(3) should be close to EA(2), the true vertical EA.

\subsection{Fundamental Gap} 

The fundamental gap ($E_{g}$) of a neutral molecule is defined as $E_{g} = \text{IP} - \text{EA}$, the difference between the vertical IP and EA of the neutral molecule. Since there 
are various ways of computing the vertical IP and EA in KS-DFT, we adopt the following three ways of calculating $E_{g}$: 
\begin{align} 
& E_{g}(1) = \text{IP}(1) - \text{EA}(1) = E_{\text{total}}(cation) + E_{\text{total}}(anion) - 2E_{\text{total}}(neutral) \label{Eg1}\\ 
& E_{g}(2) = \text{IP}(2) - \text{EA}(2) = {\epsilon}_{\text{HOMO}}(anion) - {\epsilon}_{\text{HOMO}}(neutral) \label{Eg2}\\ 
& E_{g}(3) = \text{IP}(2) - \text{EA}(3) = {\epsilon}_{\text{LUMO}}(neutral) - {\epsilon}_{\text{HOMO}}(neutral) \label{Eg3} 
\end{align} 
Note that $E_{g}(3)$ is the HOMO-LUMO gap in KS-DFT or the Kohn-Sham (KS) gap. For the exact KS-DFT, while both $E_{g}(1)$ and $E_{g}(2)$ yield the exact fundamental gap, 
there is a distinct difference between $E_{g}(2)$ and $E_{g}(3)$, due to the $\Delta_{xc}$ of the XC density functional: $E_{g}(2) - E_{g}(3) = \text{EA}(3) - \text{EA}(2) = \Delta_{xc}$. 
However, for LC hybrid functionals, as EA(3) is expected to be close to EA(2), $E_{g}(3)$ should be close to $E_{g}(2)$, the true fundamental gap \cite{SC4}.

\subsection{Optical Gap} 

The optical gap ($E_{opt}$) of a neutral molecule is defined as 
\begin{equation} 
E_{opt} = E_{\text{total}}^{\text{excited}}(neutral) - E_{\text{total}}(neutral), 
\label{Eopt} 
\end{equation} 
the energy difference between the lowest dipole-allowed excited state and the ground state, calculated at the ground-state geometry of the neutral molecule. 
Since $E_{opt}$ is an excited-state property, it cannot be directly obtained with KS-DFT. For consistency with the ground-state calculations, TDDFT 
(with the same density functionals for the ground-state calculations) is adopted to compute $E_{opt}$.

\subsection{Exciton Binding Energy} 

The exciton binding energy ($E_{b}$) of a neutral molecule is defined as $E_{b} = E_{g} - E_{opt}$, the difference between the fundamental and optical gaps. 
A system with small $E_{b}$ often possesses high charge separation efficiency, and hence is favorable for photovoltaic applications, while the opposite may be desirable for 
light-emitting devices. Therefore, it is important to study the $E_{b}$ values of $n$-PPs for understanding their potential applications. 
In this work, we adopt the following three ways of calculating $E_{b}$: 
\begin{align} 
& E_{b}(1) = E_{g}(1) - E_{opt} \label{Eb1}\\ 
& E_{b}(2) = E_{g}(2) - E_{opt} \label{Eb2}\\ 
& E_{b}(3) = E_{g}(3) - E_{opt} \label{Eb3} 
\end{align} 
For the exact KS-DFT and TDDFT, $E_{b}(1)$ and $E_{b}(2)$ yield the exact exciton binding energy, while $E_{b}(3)$ deviates from the exact exciton binding energy by an amount 
of $\Delta_{xc}$. For LC hybrid functionals, as $E_{g}(3)$ should be close to $E_{g}(2)$, $E_{b}(3)$ is expected to be close to $E_{b}(2)$, the true exciton binding energy \cite{EB}.

\section{Results and Discussion} 

\begin{figure} 
\includegraphics[scale=0.55]{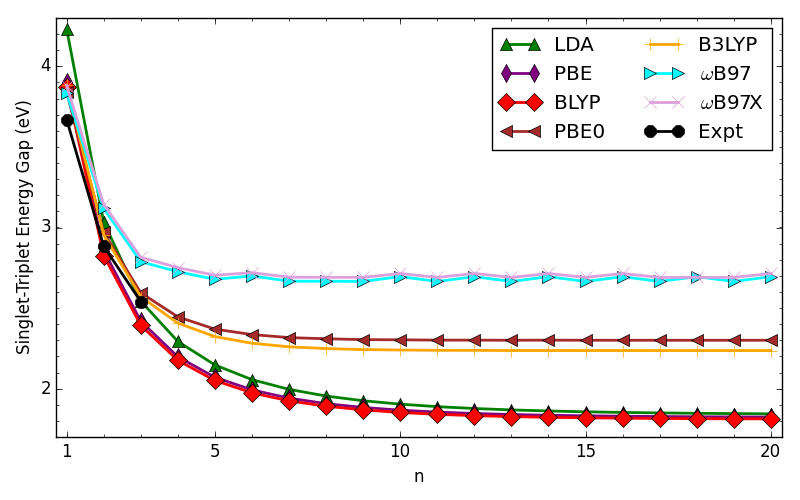} 
\caption{Singlet-triplet energy gap ($E_{\text{ST}}$) of $n$-PP as a function of the chain length, calculated using KS-DFT with various density functionals. 
Here $E_{\text{ST}}$ is calculated using Eq.\ (\ref{EST}). 
For comparison, the experimental data \cite{STgap1,STgap,conform2} are taken from the literature.} 
\label{fig:stg} 
\end{figure}

\Cref{fig:stg} shows the singlet-triplet energy gap ($E_{\text{ST}}$) of $n$-PP as a function of the chain length, calculated using KS-DFT with various XC density functionals \cite{supp}. 
The results are compared with the available experimental $E_{\text{ST}}$ data \cite{STgap1,STgap,conform2}. Overall, the calculated $E_{\text{ST}}$ curves decrease with increasing 
chain length, showing consistency with the experimental data. The ground state of $n$-PP remains singlet for all the chain lengths studied. 

Based on the calculated values of $\langle {\hat{S}}^2 \rangle$ \cite{supp}, the lowest singlet state (i.e., the ground state) of $n$-PP exhibits single-reference character (i.e., has no spin 
contamination and $\langle {\hat{S}}^2 \rangle = 0.0000$), while the lowest triplet state of $n$-PP possesses some multi-reference character (i.e., $\langle {\hat{S}}^2 \rangle > 2.0$), where 
the degree of spin contamination increases with the fraction of Hartree-Fock (HF) exchange adopted in a density functional \cite{SC1,SC2,SC3,SC4,SC5,SC6}. For $\omega$B97 and 
$\omega$B97X, the lowest triplet state of long-chain $n$-PP is slightly spin contaminated, partially degrading the accuracy of $\omega$B97 and $\omega$B97X for $E_{\text{ST}}$. 
Besides, the unphysical oscillations in the $E_{\text{ST}}$ curves obtained from $\omega$B97 and $\omega$B97X are found to be closely related to the degree of spin 
contamination \cite{supp}. 

The $E_{\text{ST}}$ value of $n$-PP at the polymer limit ($n \to \infty$) is shown in \Cref{table:expl}. The extrapolated $E_{\text{ST}}$ value is 1.67 eV for LDA and PBE, 1.65 eV for 
BLYP, 2.15 eV for PBE0, 2.08 eV for B3LYP, 2.56 eV for $\omega$B97, 2.58 eV for $\omega$B97X, and 2.05 eV for the experimental $E_{\text{ST}}$ data. Based on the calculated 
and extrapolated results, the global hybrid functionals (PBE0 and B3LYP) slightly outperform the semilocal functionals (LDA, PBE, and BLYP) and LC hybrid functionals ($\omega$B97 
and $\omega$B97X). 

\begin{figure} 
\includegraphics[scale=1.0]{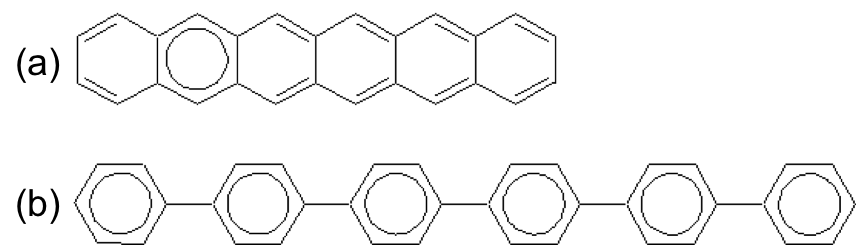} 
\caption{Kekul\'{e} structures of (a) 6-acene and (b) 6-PP. Here Clar's aromatic sextets are marked with circles \cite{Clar1}.} 
\label{fig:acene_PPP} 
\end{figure}

One may wonder why the narrowest AGNRs, i.e., $n$-PPs, possess relatively stable singlet ground states (i.e., with non-radical character), when compared to the narrowest ZGNRs, i.e., 
$n$-acenes (acenes containing $n$ linearly fused benzene rings), which have been shown to possess much less stable singlet ground states (i.e., with much smaller $E_{\text{ST}}$ values), 
and exhibit increasing polyradical character with increasing chain 
length \cite{Acene-DMRG,Acene-Jiang,TAO-LDA,GNRs-DMRG,GNRs-PHF,GNRs-MRAQCC,TAO-GGA,GNRs-MRAQCC2,GNRs-TAO}. 
We expect that the geometrical arrangements of the aromatic rings in $n$-PP and $n$-acene should be responsible for the stability of these molecules \cite{Acene-2RDM}. 
For $n \ge 3$, $n$-PP and $n$-acene are polycyclic aromatic hydrocarbons (PAHs), molecules containing three or more aromatic rings made of carbon and hydrogen atoms only. 
Based on Clar's rule, the Kekul\'{e} structure with the largest number of disjoint aromatic sextets is the most important structure for the stability of PAHs \cite{Clar1}. 
As illustrated in \Cref{fig:acene_PPP}, the aromatic rings of $n$-PP are connected with each other by a single carbon-carbon bond, isolating each aromatic ring as if $n$-PP is just the 
combination of isolated benzenes. Therefore, there are $n$ aromatic sextets in the Kekul\'{e} structure of $n$-PP. By contrast, there is only one aromatic sextet in the Kekul\'{e} structure 
of $n$-acene. Therefore, for a given number of aromatic rings $n$ ($\ge 3$), $n$-PP is always more stable than $n$-acene. 
This suggests that the geometrical arrangement of the aromatic rings of PAHs should be responsible for the properties of PAHs. 
This argument is consistent with the results of other works \cite{GNR_TDDFT,GNR_DFT1,GNR_DFT2}. 

\begin{figure} 
\includegraphics[scale=0.55]{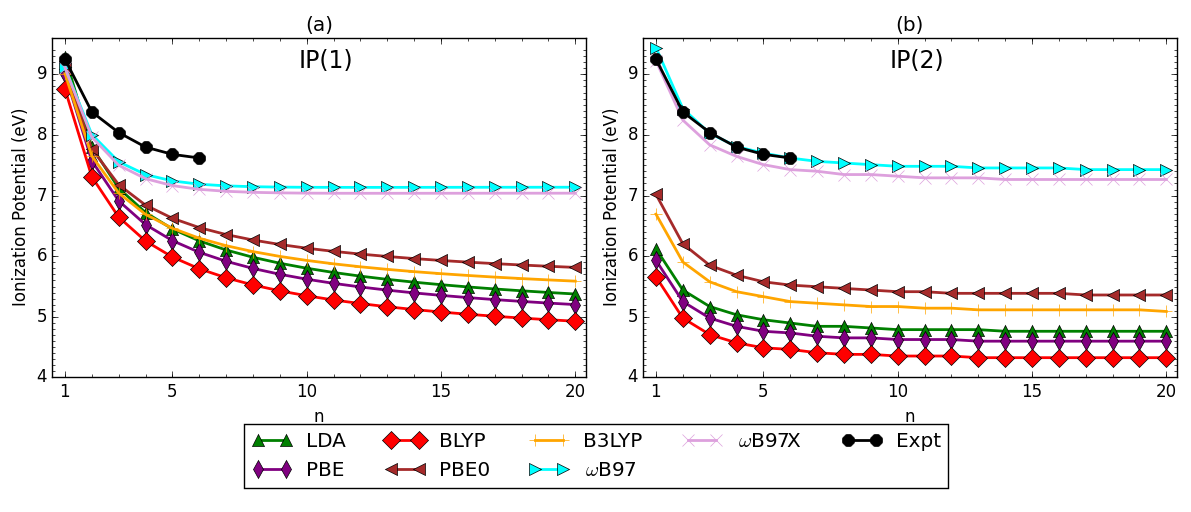} 
\caption{Vertical ionization potential (IP) for the lowest singlet state of $n$-PP as a function of the chain length, calculated using KS-DFT with various density functionals. 
Here IP(1) and IP(2) are calculated using Eqs.\ (\ref{IP1}) and (\ref{IP2}), respectively. 
For comparison, the experimental data \cite{syn2} are taken from the literature.} 
\label{fig:ip} 
\end{figure}

\begin{figure} 
\includegraphics[scale=0.55]{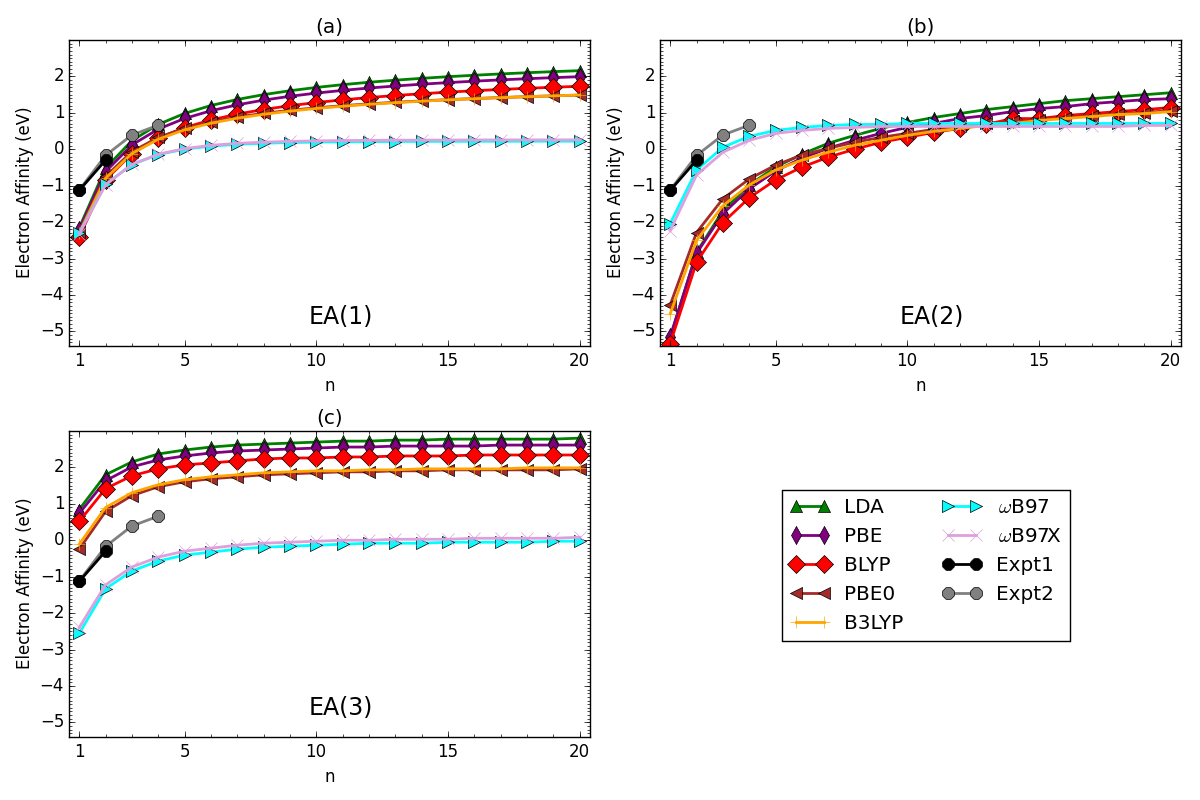} 
\caption{Vertical electron affinity (EA) for the lowest singlet state of $n$-PP as a function of the chain length, calculated using KS-DFT with various density functionals. 
Here EA(1), EA(2), and EA(3) are calculated using Eqs.\ (\ref{EA1}), (\ref{EA2}), and (\ref{EA3}), respectively. 
For comparison, the experimental data: Expt1 (vertical EA) \cite{resonance,EA1} and Expt2 (adiabatic EA) \cite{EA2,EA3}, are taken from the literature.} 
\label{fig:ea} 
\end{figure}

\begin{figure} 
\includegraphics[scale=0.55]{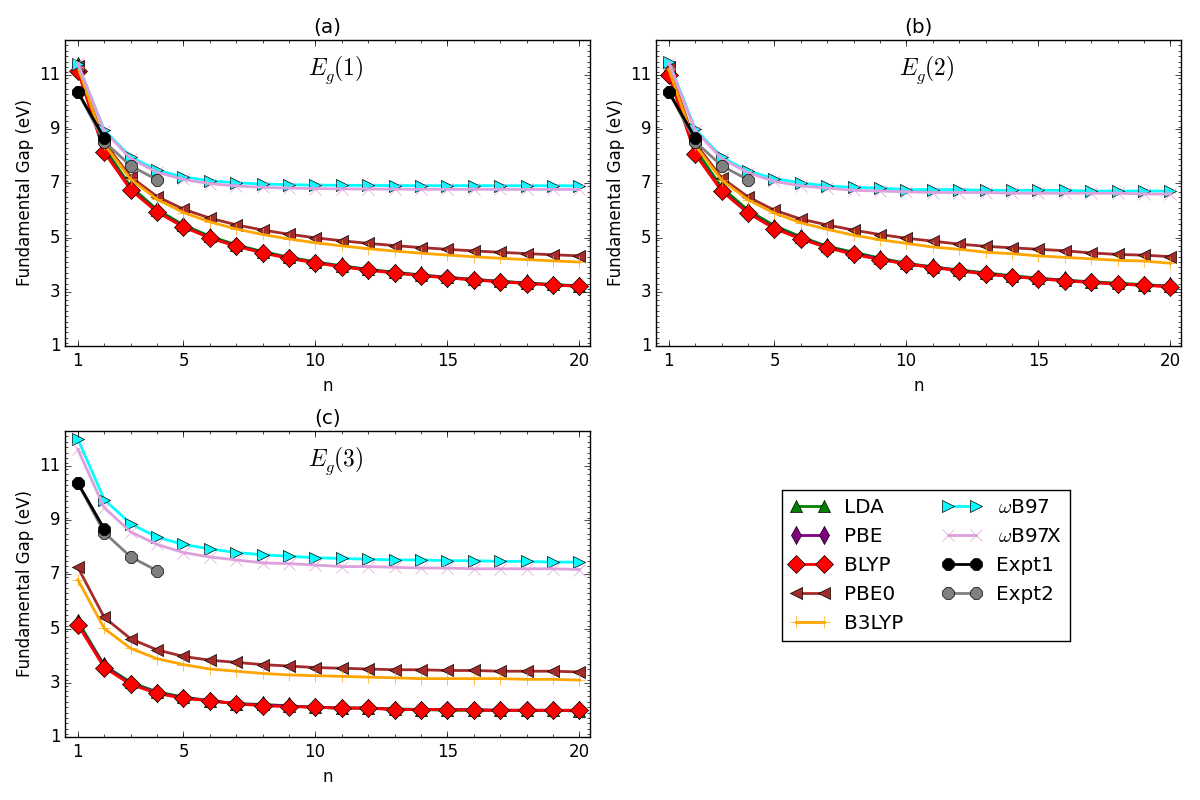} 
\caption{Fundamental gap ($E_{g}$) for the lowest singlet state of $n$-PP as a function of the chain length, calculated using KS-DFT with various density functionals. 
Here $E_{g}(1)$, $E_{g}(2)$, and $E_{g}(3)$ are calculated using Eqs.\ (\ref{Eg1}), (\ref{Eg2}), and (\ref{Eg3}), respectively. 
For comparison, the experimental data: Expt1 (vertical IP $-$ vertical EA) \cite{syn2,resonance,EA1} and Expt2 (vertical IP $-$ adiabatic EA) \cite{syn2,EA2,EA3}, are taken 
from the literature.} 
\label{fig:fg} 
\end{figure}

\begin{figure} 
\includegraphics[scale=0.55]{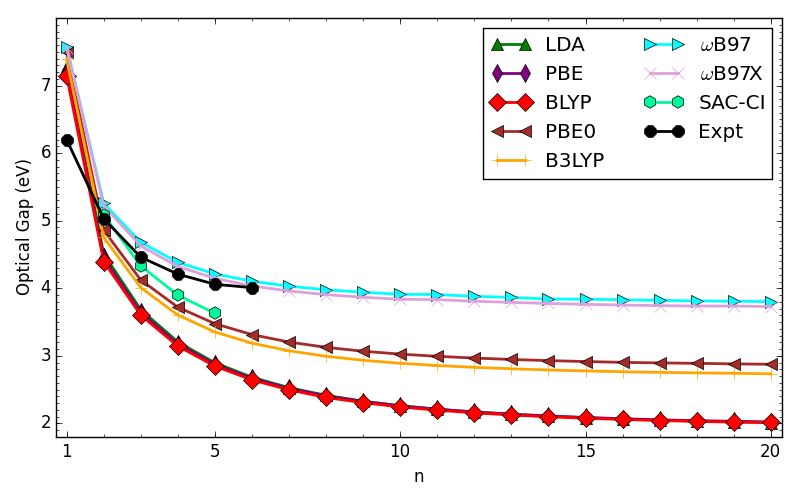} 
\caption{Optical gap ($E_{opt}$) for the lowest singlet state of $n$-PP as a function of the chain length, calculated using TDDFT with various density functionals. 
Here $E_{opt}$ is calculated using Eq.\ (\ref{Eopt}). 
For comparison, the experimental data \cite{Eopt1,Eopt3} and SAC-CI data \cite{Eopt3} are taken from the literature.} 
\label{fig:optg} 
\end{figure}

\begin{figure} 
\includegraphics[scale=0.55]{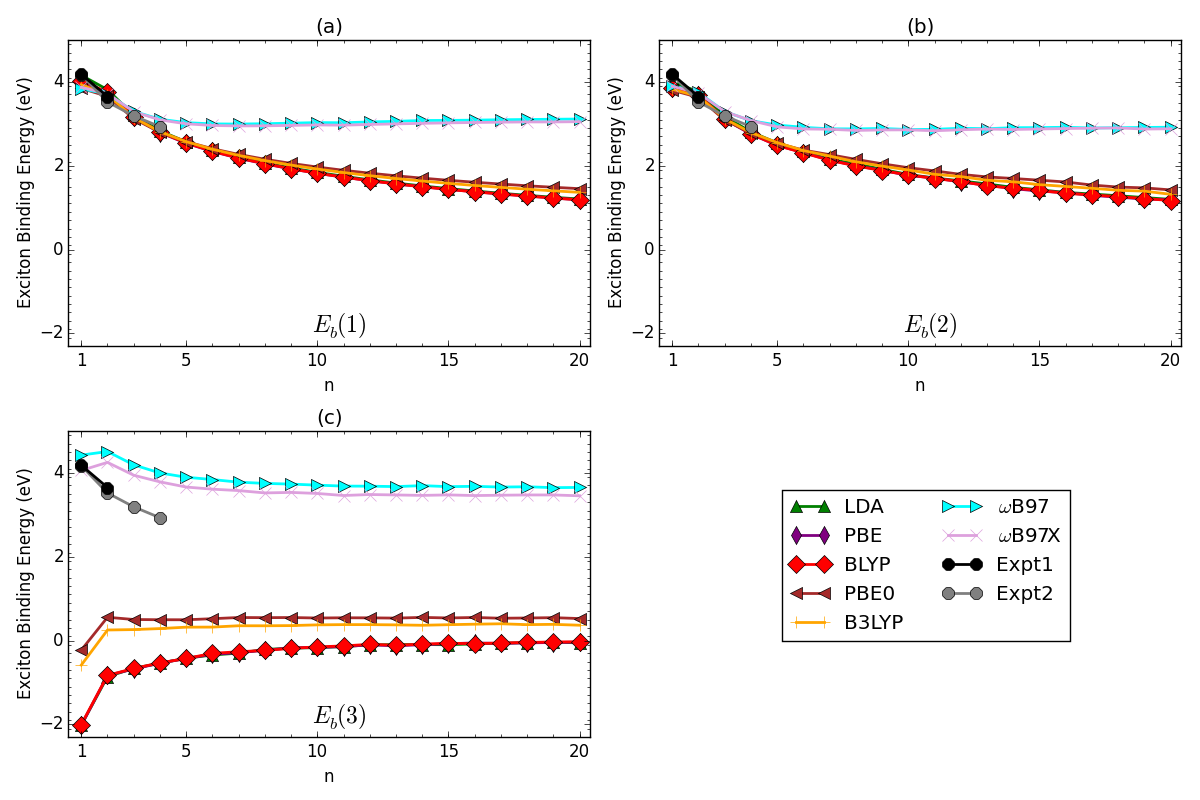} 
\caption{Exciton binding energy ($E_{b}$) for the lowest singlet state of $n$-PP as a function of the chain length, calculated using KS-DFT and TDDFT with various density functionals. 
Here $E_{b}(1)$, $E_{b}(2)$, and $E_{b}(3)$ are calculated using Eqs.\ (\ref{Eb1}), (\ref{Eb2}), and (\ref{Eb3}), respectively. 
For comparison, the experimental data: Expt1 [(vertical IP $-$ vertical EA) $-$ $E_{opt}$] \cite{syn2,resonance,EA1,Eopt1,Eopt3} and 
Expt2 [(vertical IP $-$ adiabatic EA) $-$ $E_{opt}$] \cite{syn2,EA2,EA3,Eopt1,Eopt3}, are taken from the literature.} 
\label{fig:ebe} 
\end{figure}

At the ground-state geometry of $n$-PP, the vertical ionization potential (IP) [\Cref{fig:ip}], vertical electron affinity (EA) [\Cref{fig:ea}], fundamental gap ($E_{g}$) [\Cref{fig:fg}], optical 
gap ($E_{opt}$) [\Cref{fig:optg}], and exciton binding energy ($E_{b}$) [\Cref{fig:ebe}] of $n$-PP as a function of the chain length, are calculated using KS-DFT and TDDFT with 
various XC density functionals \cite{supp}. 

As shown in \Cref{fig:ip}(a), relative to the experimental IP values \cite{syn2}, the IP(1) values calculated using $\omega$B97 and $\omega$B97X are more accurate than those 
calculated using the other functionals. The ground state of cationic $n$-PP exhibits some spin contamination ($\langle {\hat{S}}^2 \rangle > 0.75$), where the degree of spin 
contamination is vanishingly small for the semilocal functionals and global hybrid functionals, but is noticeable for the LC hybrid functionals \cite{supp}. The results are consistent 
with the argument that the larger fraction of HF exchange adopted in a density functional, the easier the resulting KS determinant becomes spin contaminated for multi-reference 
systems \cite{SC1,SC2,SC3,SC4,SC5,SC6}. 

The calculated IP(2) values [\Cref{fig:ip}(b)] are more sensitive to the choice of the XC functional than the calculated IP(1) values. The semilocal functionals and global hybrid functionals 
severely underestimate the IP(2) values due to the incorrect asymptotic behavior of the associated XC potentials. By contrast, owing to the correct $(-1/r)$ asymptote of the underlying 
XC potentials and the fact that the ground state of $n$-PP exhibits single-reference character (where $\omega$B97 and $\omega$B97X are expected to perform reasonably 
well \cite{LCAC,EB}), $\omega$B97 and $\omega$B97X yield the extremely accurate IP(2) values. As shown in \Cref{table:expl}, the IP(2) value of $n$-PP at the polymer limit is 
7.29 eV for $\omega$B97 and 7.12 eV for $\omega$B97X, which are in excellent agreement with the extrapolated experimental IP value (7.33 eV). From the calculated and extrapolated 
IP values, the IP(2) values obtained from $\omega$B97 and $\omega$B97X are reliably accurate. 

For the calculated EA(1) values [\Cref{fig:ea}(a)], due to the slight spin contamination ($\langle {\hat{S}}^2 \rangle > 0.75$) in the ground state of anionic $n$-PP, the LC hybrid functionals 
are slightly less accurate than the other functionals \cite{supp}. By contrast, as shown in \Cref{fig:ea}(b), the EA(2) values of short-chain $n$-PP ($n \le 4$) calculated using $\omega$B97 
and $\omega$B97X match very well with the experimental data: Expt1 (vertical EA) \cite{resonance,EA1} and Expt2 (adiabatic EA) \cite{EA2,EA3}, which can be attributed to the correct 
$(-1/r)$ asymptote of their XC potentials, while the other functionals significantly underestimate the EA(2) values, due to the incorrect XC potential asymptotes. However, the accuracy of 
$\omega$B97 and $\omega$B97X is slightly degraded for longer-chain $n$-PP, as the ground state of anionic $n$-PP becomes slightly spin contaminated. For the EA(3) values 
[\Cref{fig:ea}(c)], the global and LC hybrid functionals perform comparably, outperforming the semilocal functionals. From \Cref{table:expl}, the EA(2) value of $n$-PP at the polymer limit 
is 0.96 eV for $\omega$B97 and 0.90 eV for $\omega$B97X, which are in good agreement with the extrapolated experimental EA value (1.15 eV). Based on the calculated and 
extrapolated EA values, the EA(2) values obtained from $\omega$B97 and $\omega$B97X are reasonably accurate. 

For the $E_{g}(1)$ values [\Cref{fig:fg}(a)], while $\omega$B97 and $\omega$B97X slightly underestimate both the IP(1) and EA(1) values, they accurately predict $E_{g}(1)$, possibly 
due to the cancellation of errors. For the $E_{g}(2)$ values [\Cref{fig:fg}(b)], as $\omega$B97 and $\omega$B97X accurately predict both the IP(2) and EA(2) values, they accurately 
predict $E_{g}(2)$. By contrast, the semilocal functionals and global hybrid functionals severely underestimate both the $E_{g}(1)$ and $E_{g}(2)$ values. For the $E_{g}(3)$ values 
[\Cref{fig:fg}(c)], $\omega$B97 and $\omega$B97X slightly overestimate $E_{g}(3)$, whereas the other functionals significantly underestimate $E_{g}(3)$. From \Cref{table:expl}, the 
$E_{g}(1)$ value of $n$-PP at the polymer limit is 6.51 eV for $\omega$B97 and 6.36 eV for $\omega$B97X, and the $E_{g}(2)$ value of $n$-PP at the polymer limit is 6.33 eV for 
$\omega$B97 and 6.22 eV for $\omega$B97X, which are in excellent agreement with the extrapolated experimental $E_{g}$ value (6.24 eV). According to the calculated and 
extrapolated $E_{g}$ values, the $E_{g}(1)$ and $E_{g}(2)$ values calculated using $\omega$B97 and $\omega$B97X are reliably accurate. 

The optical gap ($E_{opt}$) of $n$-PP is found to be the singlet-singlet (S$_{0}$ $\rightarrow$ S$_{1}$) gap for each case studied. 
As shown in \Cref{fig:optg}, the $E_{opt}$ values calculated using $\omega$B97 and $\omega$B97X are in excellent agreement with the experimental data \cite{Eopt1,Eopt3} and 
those obtained with the highly accurate SAC-CI (symmetry-adapted-cluster configuration-interaction) method \cite{Eopt3}. Relative to the experimental data, $\omega$B97 and 
$\omega$B97X perform slightly better than the SAC-CI method. The other functionals severely underestimate the $E_{opt}$ value of long-chain $n$-PP. From \Cref{table:expl}, the 
$E_{opt}$ value of $n$-PP at the polymer limit is 3.52 eV for $\omega$B97 and 3.44 eV for $\omega$B97X, which are in excellent agreement with the extrapolated experimental 
$E_{opt}$ value (3.57 eV) and the extrapolated SAC-CI value (3.26 eV). Based on the calculated and extrapolated $E_{opt}$ values, the $E_{opt}$ values calculated using 
$\omega$B97 and $\omega$B97X are reliably accurate. 

As shown in \Cref{fig:ebe}(a) and \Cref{fig:ebe}(b), the $E_{b}(1)$ and $E_{b}(2)$ values calculated using $\omega$B97 and $\omega$B97X decrease monotonically with the increase 
of $n$, and quickly approach some constants at about $n = 5$. By contrast, the $E_{b}(1)$ and $E_{b}(2)$ values calculated using the other functionals decrease more slowly, and 
approach some constants at the larger values of $n$. For the $E_{b}(3)$ values [\Cref{fig:ebe}(c)], while $\omega$B97 and $\omega$B97X slightly overestimate $E_{b}(3)$, the other 
functionals significantly underestimate $E_{b}(3)$. Note that the $E_{b}(3)$ values obtained from the semilocal functionals are unphysically negative. From \Cref{table:expl}, the 
$E_{b}(1)$ value of $n$-PP at the polymer limit is 2.98 eV for $\omega$B97 and 2.91 eV for $\omega$B97X, and the $E_{b}(2)$ value of $n$-PP at the polymer limit is 2.80 eV for 
$\omega$B97 and 2.77 eV for $\omega$B97X, which are in excellent agreement with the extrapolated experimental $E_{b}$ value (2.63 eV). From the calculated and extrapolated 
$E_{b}$ values, the $E_{b}(1)$ and $E_{b}(2)$ values obtained from $\omega$B97 and $\omega$B97X are reliably accurate.

\section{Conclusions} 

In conclusion, we have studied the electronic and optical properties (i.e., the singlet-triplet energy gaps, vertical ionization potentials, vertical electron affinities, fundamental gaps, 
optical gaps, and exciton binding energies) of NAGNRs with different lengths, using KS-DFT and TDDFT with various XC density functionals. The ground states of NAGNRs have been 
shown to remain singlets for all the lengths studied. With the increase of the NAGNR length, the singlet-triplet energy gaps, vertical ionization potentials, fundamental gaps, optical gaps, 
and exciton binding energies decrease monotonically, whereas the vertical electron affinities increase monotonically. While the neutral NAGNRs possess stable single-reference singlet 
ground states, the lowest triplet states and the ground states of the cationic and anionic NAGNRs exhibit some multi-reference character. Nevertheless, as the degree of spin 
contamination for each case is not very severe, it seems unnecessary to employ computationally expensive multi-reference methods in this study. Overall, the electronic and optical 
properties calculated using the $\omega$B97 and $\omega$B97X functionals are in excellent agreement with the available experimental data, with the effective conjugation length of 
NAGNR being estimated to be close to 10 benzene rings. While the electronic and optical properties of NAGNRs have been shown to be controllable with the adequate choice of 
NAGNR length, how these properties vary with different widths, edge types, and edge terminations remain unanswered. We plan to address some of these questions in the near future.

\begin{acknowledgments} 

This work was supported by the Ministry of Science and Technology of Taiwan (Grant No.\ MOST104-2628-M-002-011-MY3), National Taiwan University (Grant No.\ NTU-CDP-105R7818), 
the Center for Quantum Science and Engineering at NTU (Subproject Nos.:\ NTU-ERP-105R891401 and NTU-ERP-105R891403), and the National Center for Theoretical Sciences of Taiwan. 

\end{acknowledgments}

\newpage 

\newpage 

\newpage 

\newpage 

\newpage 

\newpage 

\newpage 

\newpage 

\newpage 
\begin{table*} 
\scriptsize 
\caption{\label{table:expl} 
Singlet-triplet energy gap ($E_{\text{ST}}$), vertical ionization potential (IP), vertical electron affinity (EA), fundamental gap ($E_{g}$), optical gap ($E_{opt}$), and exciton binding 
energy ($E_{b}$) [in eV] of $n$-PP at the polymer limit ($n \to \infty$). The extrapolated values are obtained by nonlinear least-squares fittings of the corresponding properties of $1$- to 
$20$-PP, calculated using KS-DFT and TDDFT with various density functionals, where a fitting function of the form $(a + b/n)$ is adopted. For the extrapolated experimental values, only 
those with more than two data points are calculated (using the same fitting function) \cite{supp}. For each extrapolated value, the coefficient of determination $R^{2}$, which is a 
statistical measure of the goodness-of-fit ($R^{2} = 1$ for a perfect fit), is given in parenthesis. For $E_{opt}$, the extrapolated SAC-CI value is 3.26 eV ($R^{2} = 0.9757$) \cite{Eopt3}.} 
\begin{ruledtabular} 
\begin{tabular*}{\textwidth}{lcccccccc} 
& LDA & PBE & BLYP & PBE0 & B3LYP & $\omega$B97 & $\omega$B97X & Expt \\ 
\hline 
$E_{\text{ST}}$ &1.67&1.67&1.65&2.15&2.08&2.56&2.58&2.05 \\ 
&(0.9967)&(0.9954)&(0.9952)&(0.9714)&(0.9755)&(0.9289)&(0.9328)&(0.9871) \\
IP(1) &5.33&5.19&4.92&5.77&5.55&6.95&6.84&7.33 \\
&(0.9619)&(0.9598)&(0.9604)&(0.9747)&(0.9717)&(0.9762)&(0.9797)&(0.9930) \\
IP(2) &4.67&4.50&4.23&5.25&5.00&7.29&7.12&7.33 \\
&(0.9973)&(0.9970)&(0.9970)&(0.9974)&(0.9975)&(0.9972)&(0.9966)&(0.9930) \\ 
EA(1) &2.18&2.01&1.74&1.55&1.54&0.44&0.48&1.15 \\ 
&(0.9623)&(0.9618)&(0.9614)&(0.9763)&(0.9727)&(0.9893)&(0.9908)&(0.9804) \\
EA(2) &1.49&1.34&1.07&1.05&1.00&0.96&0.90&1.15 \\
&(0.9430)&(0.9415)&(0.9406)&(0.9583)&(0.9520)&(0.9835)&(0.9847)&(0.9804) \\
EA(3) &2.90&2.72&2.45&2.07&2.11&0.12&0.22&1.15 \\
&(0.9980)&(0.9974)&(0.9982)&(0.9966)&(0.9972)&(0.9974)&(0.9965)&(0.9804) \\
$E_{g}(1)$ &3.17&3.18&3.17&4.22&4.01&6.51&6.36&6.24 \\
&(0.9621)&(0.9608)&(0.9609)&(0.9755)&(0.9722)&(0.9844)&(0.9866)&(0.9882) \\
$E_{g}(2)$ &3.17&3.16&3.16&4.21&3.99&6.33&6.22&6.24 \\
&(0.9595)&(0.9597)&(0.9592)&(0.9753)&(0.9703)&(0.9914)&(0.9914)&(0.9882) \\
$E_{g}(3)$ &1.77&1.78&1.77&3.18&2.89&7.17&6.90&6.24 \\
&(0.9981)&(0.9979)&(0.9981)&(0.9975)&(0.9978)&(0.9978)&(0.9970)&(0.9882) \\
$E_{opt}$  &1.73&1.73&1.73&2.56&2.42&3.52&3.44&3.57 \\
&(0.9992)&(0.9994)&(0.9995)&(0.9976)&(0.9981)&(0.9896)&(0.9902)&(0.9944) \\
$E_{b}(1)$ &1.46&1.45&1.45&1.66&1.59&2.98&2.91&2.63 \\
&(0.8033)&(0.7852)&(0.7875)&(0.7914)&(0.7921)&(0.8118)&(0.8464)&(0.9748) \\
$E_{b}(2)$ &1.44&1.43&1.43&1.64&1.57&2.80&2.77&2.63 \\
&(0.7844)&(0.7728)&(0.7700)&(0.7872)&(0.7763)&(0.8893)&(0.8861)&(0.9748) \\
$E_{b}(3)$ &0.04&0.04&0.05&0.62&0.47&3.64&3.45&2.63 \\
&(0.9924)&(0.9904)&(0.9909)&(0.7620)&(0.8848)&(0.7947)&(0.7159)&(0.9748) \\ 
\end{tabular*} 
\end{ruledtabular} 
\end{table*} 

\end{document}